\documentclass[fleqn,usenatbib]{mnras}

\usepackage{newtxtext,newtxmath}
\usepackage{mathptmx}
\usepackage{txfonts}

\usepackage{graphicx}	
\usepackage{amsmath}	
\usepackage{amssymb}	

\usepackage{soul}

\title[Misaligned discs in black hole binaries]{Misaligned snowplough effect and the electromagnetic counterpart to black hole binary mergers}

\author[F. A. C. Pereira et al.]{
Fabr\'icia A. C. Pereira$^{1,2}$,\thanks{E-mail: fabriciaacpereira@gmail.com}
Giuseppe Lodato$^{2}$,
Irapuan Rodrigues$^{1}$,
\newauthor 
M\'arcio E. S. Alves$^{3}$ and
Daniel J. Price$^{4}$
\\
$^{1}$Universidade do Vale do Para\'iba, Av. Shishima Hifumi 2911, Urbanova 12244-000, S\~ao Jos\'e dos Campos, SP, Brazil \\
$^{2}$Dipartimento di Fisica, Universit\'a degli Studi di Milano, Via Celoria 16, 20133, Milano, Italy\\
$^{3}$Universidade Estadual Paulista (UNESP), Instituto de Ci\^encia e Tecnologia, S\~ao Jos\'e dos Campos, SP, 12247-004, Brazil\\
$^4$School of Physics \& Astronomy, Monash University, VIC 3800, Australia
}

\date{Accepted XXX. Received YYY; in original form ZZZ}

\pubyear{2018}

\begin{document}
\label{firstpage}
\pagerange{\pageref{firstpage}--\pageref{lastpage}}
\maketitle

\begin{abstract}
We estimate the accretion rates produced when a circumprimary gas disc is pushed into the primary supermassive black hole (SMBH) by the tidal force of the decaying secondary during a SMBH merger. Using the 3D Smoothed Particle Hydrodynamics (SPH) code {\sc phantom}, we extend previous investigations of co-planar discs to the case where the disc and binary orbital planes are misaligned. We consider a geometrically thin disc with inclination angles varying from 1 to 180 degrees and a binary with mass ratio $q$=$10^{-3}$. We find that discs with small inclination angles (< 10 degrees) produce an increase in luminosity exceeding the Eddington rate. By contrast, discs with inclinations between 20 and 30 degrees show a less pronounced rise in the accretion rate, whilst discs inclined by 180 degrees show no peak in the mass accretion rate. While previous analytic work predicted that the effective tidal torque drops with increasing inclination angle, we show that the misaligned snowplough effect remains important even for angles larger than the disc aspect ratio. The rise in the accretion rate produced by discs inclined at small angles to the binary orbit can produce an electromagnetic counterpart to the gravitational wave signal emitted from final stages of the binary orbital decay.
\end{abstract}




\section{Introduction}

In the centre of most galaxies lies a supermassive black hole (SMBH) with a mass exceeding $10^5M_{\odot}$ \citep{reines14}. It is believed that SMBH binaries in galactic nuclei form by successive coalescences following the merger of their respective host halos, and by accretion of gas present in the central regions of galaxies. The mechanisms assumed to form a SMBH binary after the merger of two galaxies are i) binary shrinking due to stellar or gas dynamical processes; followed by ii) coalescence due to low frequency gravitational wave emission ($\leq 10^{-3}$ Hz) \citep{mayeretal07, fillouxetal}. If the interacting galaxies are rich in gas, an accretion disc will form from gas rapidly falling towards the black holes \citep{armitage02}. 

The interaction between the SMBH binary and the surrounding material may be observable. Gas plays a key role in the binary merger process. It also powers the black hole luminosity via accretion. SMBH mergers are thus promising for multi-messenger astronomy since they produce an electromagnetic counterpart to a gravitational wave burst.

 \citet{armitage02} showed, using 1D simulations, that rapid accretion of the inner disc due to the tidal effect of the merging secondary causes an accretion rate exceeding the Eddington limit prior to the merger of the SMBH binary. By contrast, the 2D hydrodynamical simulations of \citet{baruteal12} suggested that such a `snowplough mechanism' is unlikely to work, because the binary shrinking time driven by gravitational waves is shorter than the viscous time-scale, meaning that fluid elements in the inner disc get funneled to the outer disc across the secondary gap via horseshoe orbits. However, 3D simulations carried out by \citet{ceriolietal} demonstrated the presence of a squeezing mechanism caused by the compression of the inner disc gas as the secondary companion spirals towards the more massive black hole. The resulting accretion luminosity in the final stages of the merger exceeded the Eddington rate.

Recent detections of gravitational waves have provided tests of General Relativity and direct measurements of compact objects parameters like mass and spin. These have been demonstrated in detections of gravitational radiation from stellar mass black holes binaries (events GW150914, GW151226, GW170104, GW170814 and GW170608) and by the first direct evidence of a link between a neutron star merger and short $\gamma$-ray bursts (GW170817) detected by LIGO and Virgo observatories \citep{abbott01, abbott02, abbott03}. However, the detection of gravitational waves from SMBH binaries will be possible only from future space detectors, such as LISA \citep{amaro}. 

Most previous simulations of a SMBH binary coalescence in a gaseous environment have modeled aligned disc-binary systems \citep{ceriolietal}. In the present work, we generalize the analysis to the case of misaligned disc-binary systems. Inclined binaries are expected when the black holes are spinning at an inclined angle with respect to the orbital plane. We model the disc-binary interaction with inclined discs at 1, 5, 10, 20, 30 and 180 degrees, using the 3D Smoothed Particle Hydrodynamics (SPH) code {\sc phantom} \citep{lodatoeprice10, priceefede10, price12, priceetal18}.

The paper is organized as follows: in Section~\ref{section2} we introduce the interaction process in misaligned disc-binary systems. Section~\ref{section3} describes the numerical method and initial conditions. Section~\ref{section4} shows our results varying the disc inclination angle and the corresponding accretion rates. We discuss and conclude in Section~\ref{section5}.

\section{ACCRETION DYNAMICS IN MISALIGNED DISC-BINARY SYSTEMS}
\label{section2}

Previous numerical simulations and analytic work has shown that the tidal torques of the binary secondary component can strongly influence the gas disc \citep{linpapa79, goldtre79, lubowetal15}. The dynamics of a binary embedded in a gas disc depend on the mass ratio of the binary and the gas mass. Binaries with nearly equal mass black holes have tidal torques carrying angular momentum between the gas and the binary orbit. However, for binaries with small mass ratios, the secondary tidal force can only be felt when the gas is close enough to the secondary object. Therefore, angular momentum transfer mechanism between the secondary and the gas in the accretion disc can be determined by studying the close encounters through impulsive approximation. These encounters occur on a timescale much smaller than the typical orbital time \citep{linpapa79, linpapa7902}.

The angular momentum exchange with the accretion disc can make the secondary component migrate. However, there are different types of migration depending on the relative mass of the companion. In particular, in this work we consider a low mass secondary SMBH embedded in a circumprimary disc. The low mass companion falls towards the supermassive primary. When the binary reaches separations smaller than $10^{-3}$pc the dominant process for the loss of energy and angular momentum is gravitational wave emission \citep{peters63}. The energy dissipation results in a negative torque shrinking the binary down to merger. If we assume a circular orbit, the decay rate of the binary separation due to gravitational wave emission is given by \citep{peters64}
\begin{equation}
	\left(\dfrac{da}{dt}\right)_{\rm{gw}} \simeq -\dfrac{64}{5} \dfrac{G^3 M_p^3 q_1}{c^5 a^3},
	\label{eq:decayrategw}
\end{equation}
where $M_p$ is the primary SMBH mass, $a$ is the binary separation and the $q_1$ parameter is defined as $q_1$=$q$(1+$q$), where $q$ is the binary mass ratio and the final merger time is
\begin{equation}
    \tau_{\rm{gw}} \approx \dfrac{5}{256} \dfrac{c^5 a_0^4}{G^3 M_p^3 q_1}\cdot
	\label{eq:mergertime}
\end{equation}

When the disc and binary black hole orbits are aligned, gravitational radiation emission causes the binary separation to decrease, producing an increase in luminosity that may be detectable as an electromagnetic precursor. Previous investigations for this orbital configuration yielded accretion rates of the order of $10^2$ times the Eddington limit \citep{armitage02, ceriolietal}.

\subsection{Exploring misaligned disc-binary systems}


For a SMBH binary in a circular orbit, the interaction between the misaligned disc and the binary is analogous to Lense-Thirring precession acting on an accretion disc around a single, spinning black hole \citep{lt1918, lodatoepringle06,nealonetal15}. Misaligned disc-binary interaction involves the additional effect of inclination decay due to viscous dissipation effects. For a primary black hole surrounded by an inclined disc, the gas flow is dominated by viscous torques. In this case, tilted disc-binary systems are not expected to produce an electromagnetic counterpart to the binary merger.

Binary interaction in a gaseous environment depends on the disc properties. The decoupling radius refers to the radius at which the gravitational torque becomes comparable to the viscous torque. When the initial distance between the two black holes is much larger than the decoupling radius, the binary can be surrounded by a circumbinary disc. By contrast, when the evolution is dominated by gravitational wave emission (i.e, when the black hole initial separation is small compared to the decoupling radius), then an individual disc surrounds only the more massive black hole. For a circumbinary disc, the angular momentum vector of the disc does not always coincide with that of the binary. The initial disc orientation is set by the angular momentum distribution of the gas rather than by the angular momentum of the binary \citep{hayasaki13}. 

\citet{lubowetal15} probed the effects of binary-disc inclination on Lindblad resonant tidal torques acting on a circumbinary disc. The authors studied the 2:1 inner Lindblad resonance (for m = 2) that dominates the tidal truncation of coplanar discs  by a prograde binary. For that resonance, they found a rapid decrease of the torque with inclination angle --- by a factor of about 2 for 30 degrees, by a factor of about 20 for 90 degrees and to zero for 180 degrees (counter-rotating case). 

Viscous torques can dominate Lindblad resonant torques (for m = 2) if the binary is counter-rotating. They can also dominate for smaller inclination angles, if the disc is sufficiently viscous. In this case, the gas in the disc can be captured by the secondary, flowing either into a circumbinary disc or escaping. In a inclined circumsecondary disc, the weakened tides allow mass transfer from the secondary component to the central one. Inclined discs are expected to be larger than coplanar discs due to the decrease in the resonant tidal torques \citep{lubowetal00, lubowetal15}. 

In the present work, it is assumed that the most important effect for the misalignment between the disc and the orbital plane, is that the direction of the orbital angular momentum vector does not match the spin of the primary \citep{bardeen}. Therefore, since by the Bardeen-Petterson effect the disc is aligned with the primary spin axis, this results in a misalignment with the orbital plane of the companion.

\section{NUMERICAL SIMULATIONS}
\label{section3}

We model the evolution of a misaligned disc-binary system using the {\sc phantom} smoothed particle hydrodynamics (SPH) code. {\sc phantom} solves the hydrodynamics equations in Lagrangian form using SPH \citep{price07, lodatoeprice10, priceefede10, price12, priceetal18}. The disc viscosity $\nu$ is modeled according to the \citet{ss73} $\alpha$-prescription, given by
\begin{equation}
    \nu \simeq \alpha_{\rm SS} c_{\rm s} H,
	\label{eq:sspresciption}
\end{equation}
where $\alpha_{\rm SS}$ is a dimensionless scale parameter, $c_{\rm s}$ is the sound speed of the gas in the disc and $H$ = $c_{\rm s}/\Omega$ is the disc thickness, where $\Omega$ is the Keplerian velocity. In order to model the \citet{ss73} viscosity we use the SPH artificial viscosity formalism set by the relation
\begin{equation}
    \alpha_{\rm SS} \simeq \dfrac{1}{10} \alpha^{\rm AV} \dfrac{\langle h \rangle}{H},
	\label{eq:artviscosity}
\end{equation}
where $\alpha^{\rm AV}$ is the artificial viscosity coefficient and $\langle h \rangle$ is the azimuthally averaged smoothing length. SPH employs this artificial viscosity term primarily to resolve shocks, but we use this to represent a source of viscous diffusion following \citet{lodatoeprice10}. 

The gas in the disc follows a locally isothermal equation of state with the sound speed described by the power law
\begin{equation}
    c_{\rm s} \simeq c_{{\rm s},0} R^{-\beta},
	\label{eq:sppowerlaw}
\end{equation}
where $R$ is the radial distance from the centre of mass of the binary (in cylindrical coordinates) and $c_{{\rm s},0}$ is a normalization that determine the disc thickness. The disc surface density profile is given by
\begin{equation}
    \Sigma \simeq \Sigma_0 R^{-\gamma},
	\label{eq:discsfprofile}
\end{equation}
where $\Sigma_0$ is chosen in order to set the disc mass.

In equations~(\ref{eq:sppowerlaw}) and~(\ref{eq:discsfprofile}), we chose the parameters $\beta$ = 3/2 and $\gamma$ = 3/4 in order to uniformly resolve the disc, given that the smoothing length $h \propto \rho^{-\frac13}$ (where $\rho$ is the density) is proportional to the thickness $H \propto R^{\frac34}$, so the ratio $\langle h \rangle$/$H$ in equation~(\ref{eq:artviscosity}) is constant with respect to radius.

\subsection{Initial conditions}
\label{initialcond}

 We set up a binary system of SMBHs with unequal masses on a circular orbit. The binary mass ratio is $q$ = $M_{s}$/$M_{p}$ = $10^{-3}$, where $M_p$ = $10^8 M_{\odot}$ is the mass of the primary and $M_s$ = $10^5 M_{\odot}$ is the mass of the secondary. Based on 3D SPH simulations by \citet{ceriolietal}, we chose $a_0$ = $4.75 GM_p$/$c^2$ (in code units) for the binary initial separation, corresponding to $a_0$ $\simeq$ $2.28 \cdot 10^{-5}$ pc. Moreover, since we were interested in estimating the mass accretion rate of the primary and secondary black holes, we set their accretion radii to $R_{ar,p}$ = $2 GM_p$/$c^2$ and $R_{ar,s}$ = $0.2 GM_p$/$c^2$. We assumed an initial binary orbital separation smaller than the decoupling radius ($a_{dec} \simeq 36.4 GM_p$/$c^2$) implying a circumprimary disc only. We adopt the inner disc radius $R_{\rm in}$ = $R_{{\rm ar},p}$ = $2 GM_p$/$c^2$, equal to the Schwarzschild radius and the outer radius $R_{\rm out}$ = $4.1 GM_p$/$c^2$, according to the initial separation of the black holes. Assuming the values for initial separation and accretion radii, the binary coalescence time by gravitational decay is $\tau_{\rm gw} \simeq$ 9476 $GM_p$/$c^3$ (in code units), or approximately 54 days.

We focus on inclined disc models to investigate the effects of the tidal torques by the secondary component embedded in a geometrically thin gas disc. We assume an initial disc mass of $M_{\rm disc}$ $\approx$ $1 M_{\odot}$, flat and misaligned by a specified inclination angle. The disc is truncated at an outer radius related to the intensity of the tidal torques. We neglect the outer disc, i.e. we assume it is frozen behind the companion such that the disc inner edge is located at the decoupling radius (the disc may then viscously move inwards).

We assume a disc aspect ratio $H$/$R$ = 0.01 and Shakura-Sunyaev viscosity parameter $\alpha_{\rm SS}$ = 0.01. We consider discs inclined by 1, 5, 10, 20, 30 and 180 degrees. Moreover, we also consider a control simulation of the disc without the secondary black hole. We performed seven simulations with $5 \cdot 10^5$ SPH particles for all angles mentioned above, including the simulation without no companion. The choice for the number of particles is for computational efficiency, but does not affect the quality of the numerical results as we will see in the next section. We adopted the following code units: for the length, the gravitational radius ($R_g$=$G M_p$/$c^2$); for the mass, the primary black hole mass ($M_p$) and for the time, $t$=$GM_p$/$c^3$. 

\subsection{Physical validity of the disc model in the context of AGN accretion}

The accretion rates exceeding the Eddington rate presented in this article show that the primary SMBH disc is radiatively efficient. Currently, the mechanism powering Active Galactic Nuclei (AGN) is believed to come from gas accretion onto a SMBH. The standard thin accretion disc models are normally associated with radiatively efficient AGNs. A significant fraction of the energy generated by viscous dissipation in the flow is radiated locally and the advection of energy is negligible, as a consequence the disc is geometrically thin. So that AGN discs are in general quite thin, ranging H/R $\sim 0.001-0.01$ and the bolometric luminosities correspond to about 10 per cent of the rest-mass energy of the accreting matter and approximately 40 per cent for rapidly spinning black holes \citep{ss73}.

\subsection{Disc aspect ratio}

Motivated in the investigation of electromagnetic precursors just prior to binary merger, we chose the disc aspect ratio in the inner disc based on the following reasons. \cite{tl15} estimated the fossil disc mass just prior to a SMBH merger and they found that for a $10^8 M_{\odot}$ primary black hole the inner disc mass at decoupling is of the order of $1 M_{\odot}$. This result shows that the rapid accretion of the whole circumprimary disc would produce peak luminosities of order 1--20 times Eddington luminosity. Merging binaries are expected to have thinner accretion discs and to provide an electromagnetic signature from the squeezing of the inner disc.

\cite{ceriolietal} performed simulations with thin discs for the aligned disc-binary orbital planes and they found an increase in luminosity exceeding the Eddington limit. In addition, the authors confirmed that if gas is present between the SMBHs it is squeezed and quickly accreted by the primary black hole during the orbital decay. On the other hand, if the disc is thick, gas can more easily flow outwards through the gap.

By contrast, \cite{baruteal12} found that the snowplough effect does not occur for thicker discs. Those authors showed that a small fraction (about 20 per cent) of the disc mass is accreted by the primary SMBH. This quantity does not cause any rise in the luminosity prior to binary merger and it does not excite the formation of peaks in the surface density of the disc.

\begin{figure*}
	\includegraphics[width=\columnwidth]{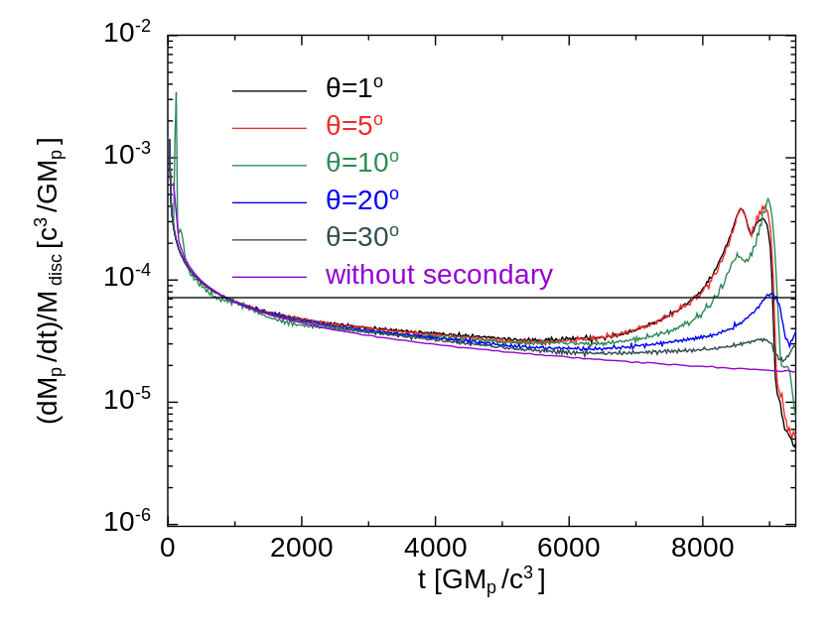} \includegraphics[width=\columnwidth]{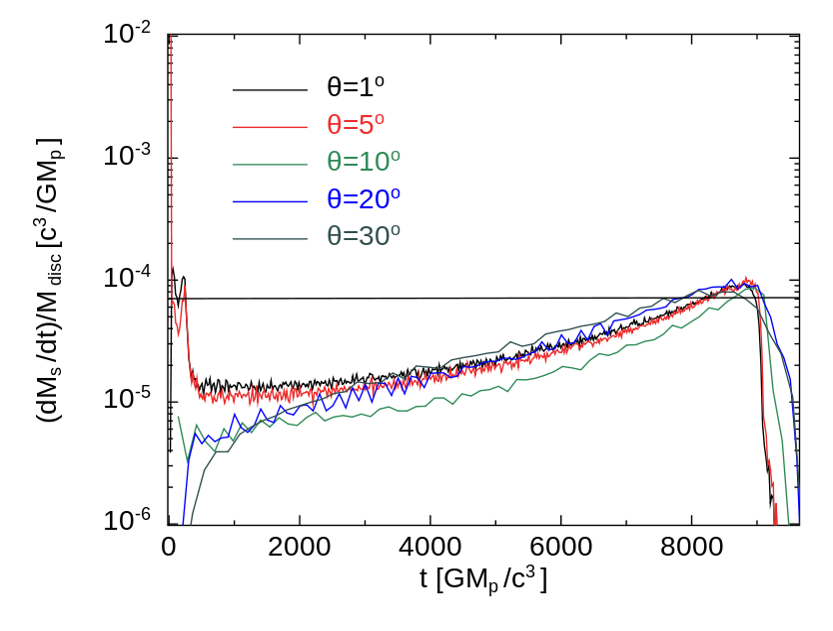}
    \caption{SMBH accretion rates in code units with discs at different inclination angles for simulations with $5 \cdot 10^5$ particles. Left panel: primary accretion rates as a function of time, show super-Eddington peaks in $\dot{M}$ for discs inclined by 1, 5 and 10 degrees. Right panel: secondary accretion rates as a function of time show squeezing phenomenon only in discs inclined by 1 and 5 degrees only. In both figures the black horizontal line shows the Eddington limit.} 
    \label{fig:allanglesaccrate}
\end{figure*}

Long before the merger of black holes embedded in thicker discs, the binary hollows out any gas present and shrinks slowly compared to the viscous timescale of a circumbinary accretion disc. So that a small fraction of the energy liberated during the merger can go into heating the gas, producing an electromagnetic afterglow \citep{miphinney}.

\section{RESULTS}
\label{section4}

 We investigate the accretion rates induced as the secondary orbit shrinks towards the primary. We compare our results with previous investigations for aligned disc-binary orbital planes.

After binary decoupling from the gas disc, the tidal torques dominate over viscous torques. At this moment, gravitational wave emission is the dominant process driving the binary to merger, with a possible electromagnetic signature in the last stages of coalescence. However, for misaligned disc-binary orbital planes, that path does not necessarily occur. Some discs are expected to be inclined, for instance when the black holes are spinning at an inclined angle with respect to angular momentum vector of the disc. 


\subsection{Varying disc inclination angles}

Figure~\ref{fig:allanglesaccrate} shows primary and secondary accretion rates as a function of time (in code units) for all simulated disc inclination angles. For discs with small inclination angles (1, 5 and 10 degrees) we found an increase in luminosity exceeding the Eddington rate and very close to the aligned disc case \citep{ceriolietal}. Table~\ref{tab:accratefinal} presents results for the angles less than 10 degrees with two times ($t_1$ and $t_2$) corresponding to the two peaks marked by episodes of strong accretion on the primary (near the binary merger). The times of the first and second spikes differ between simulations because the binary evolves embedded in a gas disc with different tilts. The double peak in the primary accretion rate occurs as the tidal torques dominate viscous torques, because the viscous time-scale is shorter than the gravitational decay time-scale. The tidal torques provide the disc angular momentum loss that drives the gas accretion onto the primary component.

\begin{figure} 
	\includegraphics[width=\columnwidth]{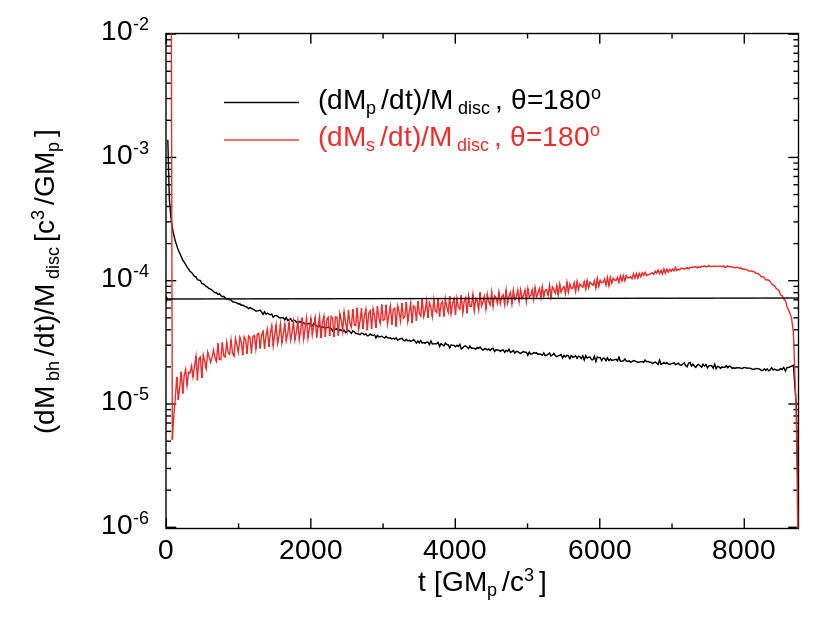}
    \caption{Primary (black line) and secondary (red line) accretion rates with circumprimary disc inclined at 180 degrees. The black horizontal line shows the Eddington limit.}
    \label{fig:allanglesprimary}
\end{figure}

The disc inclined at 10 degrees shows a less pronounced first spike in the primary accretion rate ($dM_p$/$dt$ $\simeq 10.6 M_{\odot}$/$\rm{yr}$) compared to discs with smaller angles $d{M}_p$/$dt$ $\simeq 23.1 M_{\odot}$/$\rm{yr}$ and  $d{M}_p$/$dt$ $\simeq 24.7 M_{\odot}$/$\rm{yr}$ (1 and 5 degrees, respectively). In this case, tidal torques decline due to higher inclination (10 degrees) and the primary accretion rate reaches about $d{M}_p$/$dt$ $\approx 2.35 \dot{M}_{\rm Edd}$. The inclined disc with 1 degree yielded $d{M}_p$/$dt$ $\approx 5.49 \dot{M}_{\rm Edd}$ and with 5 degrees it obtained $d{M}_p$/$dt$ $\approx 5.13 \dot{M}_{\rm Edd}$. Moreover, when the disc is inclined at 10 degrees, the evolution between first and second spikes occurs more slowly. On the other hand, the second peak on the primary accretion rate at 10 degrees shows a narrower shape, reaching in $t_2$ a accretion rate of $d{M}_p$/$dt$ $\approx 6.58 \dot{M}_{\rm Edd}$, exceeding all other accretion rates (see Table~\ref{tab:accratefinal}).

We also obtain two spikes at the initial times of binary evolution in the secondary black hole accretion rate, but these spikes occurred only in discs inclined at 1 and 5 degrees. The spikes are marked by the rapid decay of the secondary towards the primary, squeezing the inner disc gas. Our choice of a low-mass secondary ($10^5 M_{\odot}$) resulted in its interaction with the inner disc implying in an exchange of angular momentum between the secondary component and the gas.

\begin{table*}
	\centering
	\label{tab:accratefinal}
	\begin{tabular}{lccccccr} 
		\hline
		$Inclination$ & $t_1$ & $t_2$ & ($d{M}_p$/$dt$)/($M_{\odot}$/yr) [$t_{1}$] & ($d{M}_p$/$dt$)/($M_{\odot}$/yr) [$t_2$] & ($d{M}_p$/$dt$)/$M_{Edd}$ [$t_1$] & ($d{M}_p$/$dt$)/$M_{Edd}$ [$t_2$]\\
		\hline
		$\theta$=$1^o$ & 8560 & 8902 & 23.11 & 19.23 & 5.13 & 4.27\\
		$\theta$=$5^o$ & 8569 & 8930 & 24.68 & 25.80 & 5.49 & 5.73\\
        $\theta$=$10^o$ & 8513 & 8975 & 10.58 & 29.61 & 2.35 & 6.58\\
        \hline
        \end{tabular}
        \caption{Primary black hole accretion rates at final stages of the binary merger. The first column shows inclination angle of the disc, the second and the third columns show the two different times $t_1$ and $t_2$, that correspond to the first and second spikes in the accretion rate, respectively, the fourth and the fifth columns show the accretion rates in units of $M_{\odot}$/$\rm{yr}$ for each spike, the sixth and seventh columns show primary accretion rates in units of the Eddington rate ($d{M}_{Edd}$/$dt$)=$4.5 M_{\odot}$/$\rm{yr}$ for both times.}
\end{table*}

Figure~\ref{fig:allanglesaccrate} also shows the accretion rates with discs tilted at 20 and 30 degrees. However, discs with inclinations between these two angles showed a much less pronounced rise in the mass accretion rate, because the snowplough effect is important for misalignment up to 10 degrees, much larger than $H$/$R$. In particular, for 20 
degrees of inclination we find only the first spike, with a primary accretion rate of order $1.1\dot{M}_{\rm Edd}$. Previous work by \citet{lubowetal15} predicted such results based on analytic calculations of tidal torques acting on misaligned discs. By contrast, the primary accretion rate without the secondary presented in Figure~\ref{fig:allanglesaccrate} shows no peak in the accretion rate due to the absence of the companion black hole and of the snowplough mechanism.

Figure~\ref{fig:allanglesprimary} reports the accretion rates in a circumprimary disc initially inclined at 180 degrees. No peaks are observed in the primary and secondary accretion rates. For a counter-rotating disc the tidal torques decline to zero, because viscous torques dominate over binary gravitational torques. These results are also in accordance with the analytic work by \citet{lubowetal15}.

\begin{figure} 
	\includegraphics[width=\columnwidth]{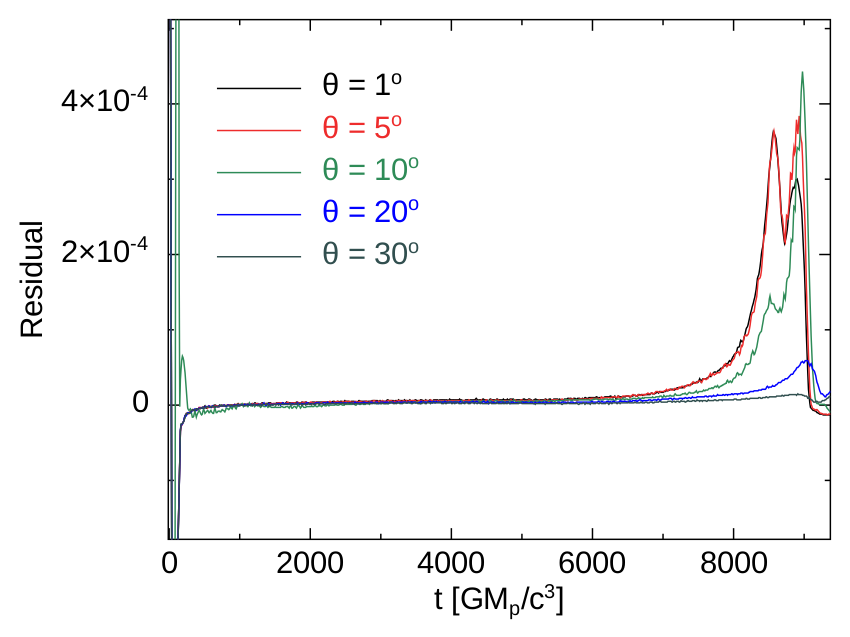}
    \caption{Residual between the accretion rates of the primary SMBH with and without the secondary.}
    \label{fig:residual}
\end{figure}

Figure \ref{fig:residual} shows the residual between the accretion rates of the primary with and without the companion. This figure clearly demonstrate what is the amount of the accretion rate due the presence of the secondary SMBH and of the snowplough effect. When comparing the accretion rate of the primary with and without the secondary, it can be seen that they are equal until $t \sim 5000$ (in code units). After this time, the snowplough effect for small inclination angles becomes important and the accretion rate increases with respect to the accretion rate without the companion.

\begin{figure*}
	\includegraphics[width=\textwidth]{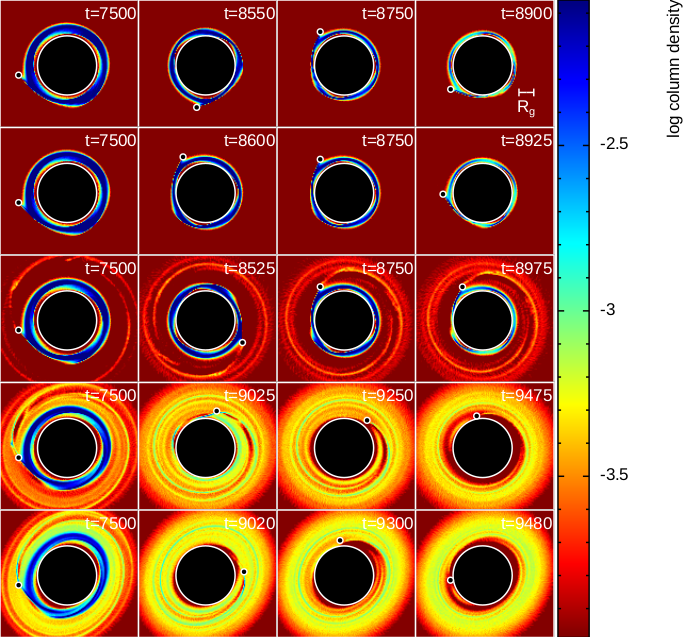}
    \caption{Snapshots of column density (logarithmic scale). Each row indicates the evolution of a misaligned binary-disc system at 1, 5, 10, 20 and 30 degrees, respectively. The supermassive black holes are represented by black filled circles with sizes  corresponding to the primary ($R_{ar,p}$=2 $GM_p$/$c^2$) and secondary ($R_{ar,s}$=0.2 $GM_p$/$c^2$) accretion radius. Time (in code units) is shown in the top right corner. The length unit is the gravitational radius $R_g$=$G M_p$/$ c^2$, where $M_p$=$10^8 M_{\odot}$ is the primary black hole mass.}
    \label{fig:allanglessnapshots}
\end{figure*}

Figure~\ref{fig:allanglessnapshots} shows snapshots of column density in the $xy$ plane. In this plot each row presents the misaligned disc-binary evolution with discs inclined at 1, 5, 10, 20 and 30 degrees, from top to bottom. The second and fourth columns show the times corresponding to the first and second spikes in the primary accretion rate. In particular, the disc inclined at 20 degrees shows only a more pronounced first peak at time $t$=9025, reaching a primary accretion rate of the order of the Eddington rate. Discs with inclination angles at 1, 5 and 10 degrees produce an increase in luminosity exceeding the Eddington rate. The snapshots shown in Figure~\ref{fig:allanglessnapshots}, with misaligned discs with $\theta >10$ degrees show more extended discs compared those inclined with angles smaller than 5 degrees.

Figure \ref{fig:allangle180snapshots} shows the evolution for a counter-rotating disc ($\theta$=180 degrees) showing a gap where the gas moves without falling towards the central black hole. In this case, the final merger take place in vacuum \citep{miphinney}.  

\subsection{Effect of thicker discs}

The disc thickness can affect the amount of gas ejected outside the orbit of the secondary black hole. In order to investigate if the thicker discs just prior to the binary merger show electromagnetic evidences, we performed simulations increasing the disc aspect ratio. In these simulations, we used inclination angles with 1 and 5 degrees, H/R=0.05, $5 \cdot 10^5$ SPH particles and the same initial conditions shown in subsection \ref{initialcond}.

Figure \ref{fig:thickerdisc} shows the accretion rates of the primary and secondary black holes for a thicker disc (H/R=0.05), with different inclination angles, normalized by initial disc mass. In both figures, we do not see any effects of the forced compression, independent of the disc inclination. All the mass is swallowed before it could be squeezed by the secondary black hole. These results are in agreement with previous simulations performed by \cite{ceriolietal} for the same value of disc aspect ratio.

\subsection{Numerical resolution}

Under the SPH formalism the spatial resolution of a simulation increases in denser regions due to the Lagrangian formulation. The choice of initial conditions is largely regulated by the available computational resources. Here, all our simulations used $5 \cdot 10^5$ SPH particles. Using higher resolution does not change the appearance of the peaks in accretion rates, but it shows improved values for the rates. Previous hydrodynamical simulations performed by \citet{ceriolietal}, for the case of an aligned disc, examined different numbers of SPH particles and resolution effects. The authors emphasized that the two spikes in primary accretion rate are present at different resolutions ($5 \cdot 10^5$, $1 \cdot 10^6$ and $2 \cdot 10^6$ SPH particles), indicating this is not a numerical artefact. However, the simulation with $2 \cdot 10^6$ particles had the most significant enhancement in the primary accretion rate, with 4.5 times better the accretion spike compared to a resolution similar to ours. But, as seen in Figures \ref{fig:allanglesaccrate} and \ref{fig:allanglesprimary}, the low resolution did not affect our overall conclusions.

\begin{figure*}
	\includegraphics[width=\textwidth]{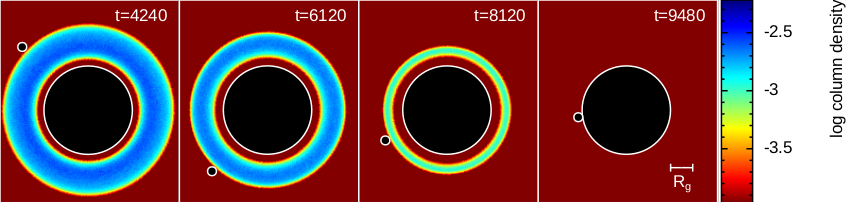}
    \caption{Snapshots of column density during the evolution of a circumprimary disc inclined at 180 degrees. The code unit for length assumed is the gravitational radius $R_g$=$G M_p$/$c^2$.}
    \label{fig:allangle180snapshots}
\end{figure*}

\section{CONCLUSIONS}
\label{section5}

We performed SPH simulations for misaligned disc-binary systems, varying the disc inclination angle, in order to investigate a possible electromagnetic precursor during the final inspirals of a binary system of SMBHs. We concentrated on a model with a gas disc surrounding the primary black hole, while the secondary black hole spirals towards the primary. We assumed a binary mass ratio $q$=$10^{-3}$, a thin accretion disc with aspect ratio $H$/$R$=0.01 and an inner disc mass of the order of $\approx 1 M_{\odot}$. We chose the disc aspect ratio in the inner disc based on the previous work of \citet{tl15}. Those authors estimated the fossil disc mass just prior to a SMBH merger and they found that for a $10^8 M_{\odot}$ primary black hole the inner disc mass at decoupling is of the order of $1 M_{\odot}$. Thus merging binaries are expected to have thinner discs.

We found that inclined discs with inclination angles of 1, 5 and 10 degrees lead to the mass accretion rates exceeding the Eddington rate between the times $t \simeq 8500 - 8975$, e.g., near binary merger time $\tau_{gw} \simeq 9476$. By contrast, inclined circumbinary discs between 20 and 30 degrees showed a much less pronounced rise in the accretion rates (see Figure \ref{fig:allanglesaccrate}). On the other hand, discs with inclination at 180 degrees showed no increase in the accretion rates (see Figure \ref{fig:allanglesprimary}). The inner mass of the disc is rapidly accreted just before of the merger, leading to an abrupt increase in the mass accretion rate above the Eddington limit. However, for inclination larger than 10 degrees the accretion rate fell abruptly with primary accretion rates smaller than the Eddington rate. These results show that the snowplough effect is important for misalignment up to 10 degrees, much larger than $H$/$R$.

Moreover, secondary accretion rates for inclined discs at 1 and 5 degrees presented double peaks confirming the squeezing phenomenon at initial times of binary evolution. The companion black hole quickly migrates towards the central black hole sweeping up the gas in the inner disc. Angles larger than 5 degrees did not present any pronounced rise in the secondary accretion rates during the binary evolution (see right panel from Figure \ref{fig:allanglesaccrate}) and the misaligned discs show somewhat larger than the those with inclinations smaller than 5 degrees (see Figure \ref{fig:allanglessnapshots}). 

Our results agree with previous analytic works by \citet{lubowetal15}, who obtained a decrease in resonant tidal torques with increasing inclination angle; we also compared the results of the electromagnetic precursors studied by \citet{ceriolietal} and found an increase in luminosity for small inclination angles (< 10 degrees), exceeding the Eddington rate. Furthermore, from the simulations we were able to identify the presence of the squeezing mechanism (for 1 and 5 degrees) during the secondary migration, such as found by \citet{ceriolietal} for aligned disc-binary systems. It should be noted, however, that we performed low resolution simulations ($N_{\rm part}$ = $5 \cdot 10^5$), while \citet{ceriolietal} considered different numerical resolutions. In particular, for a high resolution simulation ($N_{\rm part}$ = $2 \cdot 10^6$) they obtained an improvement in the accretion rates of the order of about 10 times the Eddington rate when compared with our results.

We also performed simulations with counter-rotating discs, which we found to be in concordance with the results of \citet{lubowetal15}. Those authors found that the tidal torque is zero, because viscous torques dominate over the resonances. In this case, since the disc is inclined at 180 degrees, there is an angular momentum cancellation leading to direct gas accretion on a dynamical time-scale, since in this case the gas only needs to move into the created gap and not directly on to the primary black hole. The final black hole merger thus occurs in vacuum \citep{miphinney}, as shown in Figure \ref{fig:allangle180snapshots}.

\begin{figure*}
	\includegraphics[width=\columnwidth]{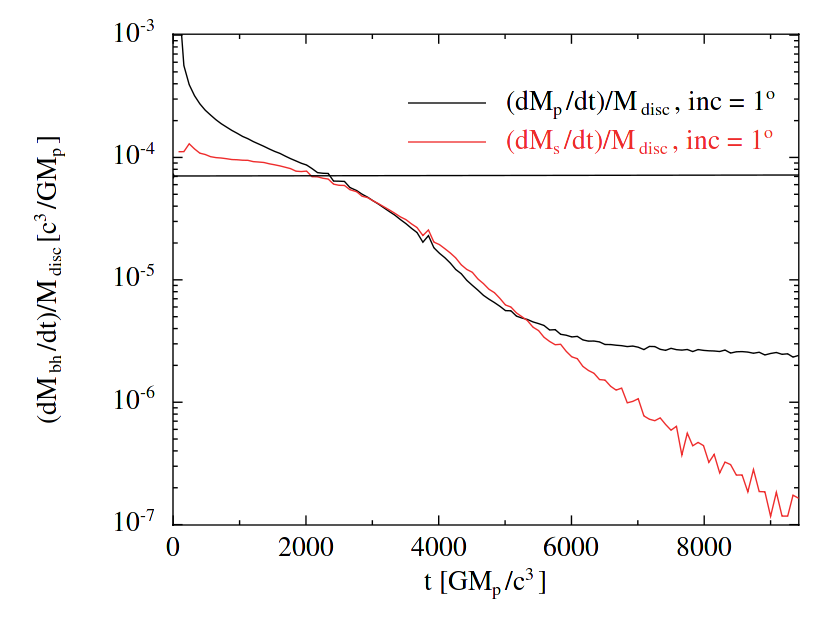} \includegraphics[width=\columnwidth]{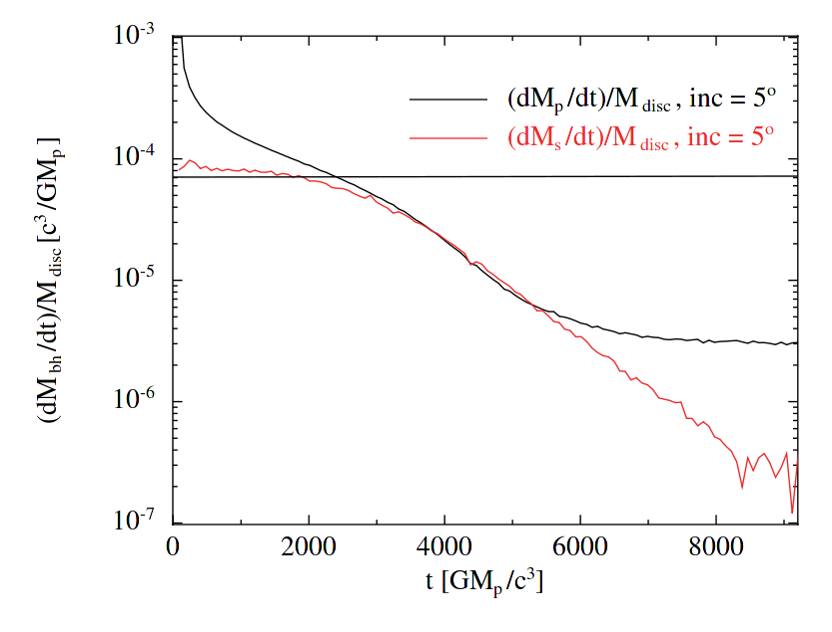}
    \caption{SMBH accretion rates in code units for a thicker disc with disc aspect ratio H/R=0.05. Left panel: primary and secondary accretion rates as a function of time for a disc inclined at 1 degree. Right panel: primary and secondary accretion rates for a disc inclined at 5 degrees. The black horizontal line shows the Eddington limit.}
    \label{fig:thickerdisc}
\end{figure*}

We ran simulations with a thicker disc (H/R = 0.05) motivated in the investigation of electromagnetic signatures just prior to binary merger. We found no effect of the forced compression, independent of the disc inclination (see Figure \ref{fig:thickerdisc}). These results are in agreement with previous work of \citet{baruteal12}. Those authors found that the snowplough effect does not occur for thicker discs. In order that a small fraction of the energy liberated during the merger can go into heating the gas, producing an electromagnetic afterglow \citep{miphinney}.

Although we have investigated tidal transport of a small mass ratio $q$=$10^{-3}$, with the secondary component migrating via gravitational waves emission towards the primary, in low resolution simulations, our results are in agreement with earlier studies. A wider investigation at higher resolution may increase the predicted mass accretion rates for small inclination angle values. 

We also concluded that our results can be applied to electromagnetic counterparts from stellar mass black hole mergers. In that case, if even a fraction of gas remains on the disc, the accretion rates produced when the gas disc is pushed into the primary black hole by the decaying secondary during the merger, lead to accretion peaks comparable or exceeding to Eddington rate. The electromagnetic signature can occur about 8-10 days before the binary final merger. 

A scenario where mergers of stellar mass black hole binaries driven by gravitational radiation producing an electromagnetic signature had already been suggested by \citet{mink17}. Those authors argued that scenario happens if a low mass accretion disc survives until coalescence. Moreover, they proposed that the disc responds to some processes of the final evolutionary phases (such as sudden mass loss and gravitational wave recoil) within hours of the merger. The electromagnetic signal will possibly arise in medium energy X-rays, perhaps extending to the infrared and last at least a few hours. In addition, \citet{martinetal18} investigated the evolution of a disc around a merging stellar mass black hole binary in the two extreme limits of an accretion disc and a decretion disc. They concluded that dynamical readjustment of the disc after the black holes merger is probably to release significant energy in electromagnetic form, corroborating with previous results by \citet{mink17}.

In conclusion, the 3D SPH simulations performed in this work predict the existence of electromagnetic precursors from the primary and from the secondary in a binary system of SMBHs when the angle of misalignment of the circumprimary accretion disc is small (less than 10 degrees), though we have only investigated a limited parameter space. This work suggests a link between electromagnetic signals and the gravitational radiation potentially detected by ground-based and future space detectors.

\section*{Acknowledgements}

FACP acknowledges financial support from CAPES (88881.133179/2016-01). 
MESA would like to thank the Brazilian agency FAPESP for financial support (grant 13/26258-4). DJP acknowledges financial support from the Australian Research Council (FT130100034). All figures in this paper were produced using SPLASH \citep{price07}. 



\bibliographystyle{mnras}
\bibliography{snowplough} 

\bsp	
\label{lastpage}
\end{document}